\documentclass[10pt,twocolumn,letterpaper]{article}

\usepackage{cvpr}
\usepackage{times}
\usepackage{epsfig}
\usepackage{graphicx}
\usepackage{amsmath}
\usepackage{amssymb}

\usepackage{times}
\usepackage{url}
\usepackage{helvet}
\usepackage{courier}
\usepackage{graphicx}
\usepackage{multirow}
\usepackage{amssymb}
\graphicspath{{./}{./Fig/}{./Figs/}}
\usepackage{amsmath,amsthm}
\usepackage{booktabs}
\usepackage{mathrsfs}
\usepackage{epstopdf}

\usepackage{graphics}
\usepackage{epstopdf}
\usepackage{amssymb}
\usepackage{subfig}
\usepackage{algorithm}
\usepackage{algorithmic}
\usepackage{tabularx}
\usepackage{caption}
\usepackage{extarrows}
\newcommand{\eat}[1]{}
\usepackage{verbatim}
\usepackage{latexsym}

\usepackage{times}
\usepackage{url}
\usepackage{helvet}
\usepackage{courier}
\usepackage{graphicx}
\usepackage{multirow}
\usepackage{amssymb}

\cvprfinalcopy 

\def\cvprPaperID{****} 

\renewcommand{\paragraph}{\textbf}
\expandafter\def\expandafter\normalsize\expandafter{%
	\normalsize\setlength\abovedisplayskip{3pt}}
\expandafter\def\expandafter\normalsize\expandafter{%
	\normalsize\setlength\belowdisplayskip{3pt}}

\begin{document}

\title{Binary Subspace Coding for Query-by-Image Video Retrieval}

\author{Ruicong Xu$^\dag$, Yang Yang$^\dag$, Yadan Luo$^\dag$, Fumin Shen$^\dag$, Zi Huang$^\ddag$ and Heng Tao Shen$^\dag$\\
$^\dag$University of Electronic Science and Technology of China\\
$^\ddag$The University of Queensland\\
{\tt\small \{ranranxu95,dlyyang,lyadanluol,fumin.shen\}@gmail.com, huang@itee.uq.edu.au, shenhengtao@hotmail.com}
}

\maketitle

\begin{abstract}
   The query-by-image video retrieval (QBIVR) task has been attracting considerable research attention recently. However, most existing methods represent a video by either aggregating or projecting all its frames into a single datum point, which may easily cause severe information loss. In this paper, we propose an efficient QBIVR framework to enable an effective and efficient video search with image query. We first define a similarity-preserving distance metric between an image and its orthogonal projection in the subspace of the video, which can be equivalently transformed to a Maximum Inner Product Search (MIPS) problem.
   Besides, to boost the efficiency of solving the MIPS problem, we propose two asymmetric hashing schemes, which bridge the domain gap of images and videos. The first approach, termed Inner-product Binary Coding (IBC), preserves the inner relationships of images and videos in a common Hamming space. To further improve the retrieval efficiency, we devise a Bilinear Binary Coding (BBC) approach, which employs compact bilinear projections instead of a single large projection matrix. Extensive experiments have been conducted on four real-world video datasets to verify the effectiveness of our proposed approaches as compared to the state-of-the-arts.
\end{abstract}


\section{Introduction}
In recent years, we have witnessed a tremendous explosion of multimedia data (e.g., images, videos) on the Web driven by the advance of digital camera, high-speed Internet, massive storage, etc. Among all the types of multimedia data, videos have been playing a significant role in reshaping the ways of recording daily life, self-expression and communication~\cite{Yu2016Web}.
And video retrieval has drawn considerable research attention by its extensive application value in the social networking, such as query-by-image video retrieval (QBIVR), which is applied to a variety of real-life applications, ranging from searching video lectures using a slide, to recommending relevant videos based on images, to searching news videos using photos~\cite{TAVR15}. Hashing~\cite{shen2016fast,yang2015robust} has been proposed to efficiently solving the problem of multimedia retrieval.

However, the QBIVR task is challenged by the similarity-preserving measurement of images and videos and an efficient retrieval method for the huge dataset. For superior similarity-preserving, it is closely connected to feature representation of videos~\cite{Yi2016Bidirectional}. Considerable research endeavors~\cite{Araujo2015Efficient,PerronninSM10} have been dedicated to developing effective schemes for improving the global signature of the whole video, and promising methods
 in subspace representations such as single or mixture of linear subspaces~\cite{WangC09a}, affine subspaces~\cite{HuMO11}, covariance matrix~\cite{VemulapalliPC13} have demonstrated their superiorities in underpinning a range of multimedia applications. Subspace representations of videos completely preserve the rich structural properties of visual objects such as viewpoint, location, spatial transformation, movement, etc, which superior to a single point representation in a high-dimensional feature space. However, they also make the similarity-preserving measurement between subspace and point data even harder, one existing method is to compute the similarity between the query image and each frames of the video and then integrate these similarities by averaging or taking the maximum. Obviously, this measurement suffers from high computational cost and massive storage, as well as ignores the correlations among the video frames.

 Meanwhile, to achieve large-scale fast retrieval, several powerful hashing methods have been proposed. However, they are unsatisfied for the QBIVR task because of their non-trivial due to the different modalities of videos and images. By projecting each video (i.e.,subspace) into a datum point in a certain high-dimensional space, Basri et al.~\cite{ANS2011} proposed an Approximate Nearest Subspace algorithm to solve the QBIVR task. Then the point-to-subspace search problem is reduced to the well-known point search problem which can be addressed by approximate nearest neighbor (ANN) techniques. Nonetheless, the performance is far from ideal effect due to the inevitable loss of physical and/or geometric structural information in inter-frame structure and intra-frame relationships resulting from the operation of aggregation and/or projection of video frames. So it is urged to propose an effective strategy to define the similarity of two distinct modalities of query point and subspace database and  exert ideal search efficiency.

\begin{figure*}[t]
  \centering
  \includegraphics[width=.95\linewidth]{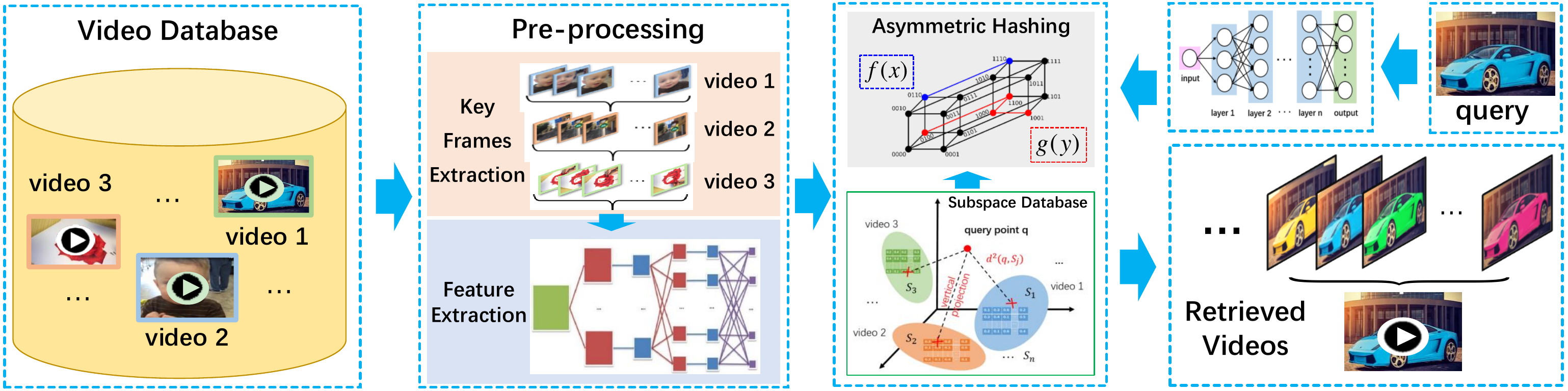}\\
  \caption{The flowchart of BSC framework. We first represent videos as subspaces of frames, and then asymmetrically project images and videos into a common Hamming space, where we can efficiently retrieve the most relevant videos provided with an image query.
  }
  \label{fig:framework}
\end{figure*}

To tackle the aforementioned two challenges in QBIVR task, we propose a novel retrieval framework, termed Binary Subspace Coding (BSC), which can fully explore the genuine geometric relationships between query point (image) and subspace (video) as well as provide significant efficiency improvements. In particular, we measure image-to-video similarity by calculating the $\ell_2$ distance of the image and its orthogonal projection in the subspace of the video and then equivalently transform the target to a Maximum Inner Product Search (MIPS) problem. To further accelerate search process and simplify the optimization, our BSC framework employs asymmetric learning strategy to generate different hash codes/functions for images and videos, which can narrow their domain gap in the common Hamming space and suitable for high-dimensional computation. Specifically, two asymmetric hashing models are designed. We first propose an Inner-product Binary Coding (IBC) approach which preserves image-to-video inner-relationships and guarantees high-quality binary codes by a tractable discrete optimization method. Moreover, we also devise a Bilinear Binary Coding (BBC) approach to significantly lower the computational cost by exploiting compact bilinear projections instead of a single large projection matrix.

We illustrate the flowchart of our proposed BSC framework in Figure~\ref{fig:framework}. The main contributions of our work are summarized as follows:
\begin{itemize}
	\item We devise a novel query-by-image video retrieval framework, termed \emph{Binary Subspace Coding} (BSC). We define an image-to-video distance metric to preferably preserve geometric information of data.
	\item We propose two asymmetric hashing schemes to unify images and videos in a common Hashing space, where the domain gap is minimized and efficient retrieval is fully supported.
 \item Extensive experiments on four datasets, i.e., BBT1, UQE50, FCVID and a micro video dataset collected by ourselves demonstrate the effectiveness of our approaches.
\end{itemize}

The reminder of this paper is organized as follows. Section 2 gives an introduction to the related work. In Section 3, we elaborate the details of our BSC framework. Section 4 demonstrates the experimental results and analysis, followed by the conclusion of this work in Section 5.

\section{Related Work}
In this section, we give a brief view of previous literatures that are closely related to our work in some aspects.

Recently, there has been a significant research interest in video retrieval such as event detection~\cite{yan2015event,gan2015devnet} and recommendation system~\cite{jiang2013understanding}. One type of video retrieval, the QBIVR task, is urgently in a need of performance boosting for its widespread applications and an effective informative representation of video undoubtly needs to be researched as an guarantee for retrieval quality~\cite{Hong2015Learning}. Typically, shots indexing, sets aggregation and global signature representation are three prevailing video retrieval methods. Differ from the severe scarcity of intra-frame relationships and sensitivity of opting training sets respectively in the first two patterns, global signature representation methods could reserve pretty rich structure information, such as inter-frame and intra-frame relationships which contributes to modelling pertinently and accurately. However, high computational costs and efficient retrieval speed is hard to handle, even though Araujo proposed~\cite{araujo2015temporal}
an integral representation method which reduces retrieval latency and memory requirements by Scalable Compressed Fisher Vectors (SCFV), but sacrifices a plethora of original spatial-temporal features which are crucial for integral representations. At the same time, a suitable similarity-preserving measurement between images and videos also merits attention which results in more difficulities in the QBIVR task.

By preserving invariant domain with low-dimensional structure information and then projecting affinity matrix into a datum point, an easier similarity-preserving measurement between images and videos was proposed for the QBIVR task by using approximate nearest neighbors (ANN) search methods~\cite{ANS2011}. Meanwhile, several powerful hashing ~\cite{Shen2016Semi} ,i.e., supervised hashing, semi-supervised hashing, unsupervised hashing did bring light to ANN search problem for pursuing efficiency. Although supervised hashing methods have demonstrated promising performance in some applications with semantic labels, it's troublesome or even impossible to get semantic labels in many real-life applications. Besides, the learning process is by far more complex and time-consuming than unsupervised techniques especially dealing with high-resolution videos or images. Some classical unsupervised methods include Spectral Hashing (SH)~\cite{SH08}, preserving the Euclidean distance in the database; Inductive Manifold Hashing (IMH)~\cite{Song2013Inter}, adopting manifold learning techniques to better model the intrinsic structure embedded in the feature space; Iterative Quantization(ITQ)~\cite{TPAMI.2012.193}, focusing on minimizing quantization error during unsupervised training. Other noticeable unsupervised hashing methods including anchor graph hashing (AGH)~\cite{AGH2011} and scalable graph hashing with feature transformation (SGH)~\cite{SGH15} directly exploit the similarity to guide the hashing code learning procedure and achieve a good performance.

Although the above hashing methods can efficiently deal with complexity of computational cost and storage, the different modalities of images and videos are neglected which can cause domain gap between images and videos. Some work have already focused on the domain gap. One solution proposed by Yan~\cite{LiWHSC15}, dubbed Hashing across Euclidean Space and Riemannian Manifold (HER), learns hash functions in a max-margin framework across Euclidean space and Riemannian manifold, but becomes unsuitable for large-scale database owing to unaffordable time when dimension grows. Shirvastava and Li~\cite{Shrivastava015} also proposed an Asymmetric Locality-Sensitive Hashing (ALSH) which performs a simple asymmetric transformation on data pairs for different learning. Inspired by their work and a dimensionality reduction bilinear projections method by Gong~\cite{GongKRL13}, we aim to seek a more powerful asymmetric binary learning framework which properly balances QBIVR task and high quality hashing based on subspace learning.

\section{Binary Subspace Coding}
In this section, we describe the details of our BSC framework in two different schemes. We first present the geometry-preserving distance metric between images and videos and deduce how the objective is transformed to MIPS problem. Then,
we respectively introduce the two different asymmetric learning approaches of hash codes/fucntions which perfectly solve domain gap between images and videos.

\subsection{Problem Formulation}
Given a database of $k$ videos, denoted as $\mathcal{S}=\{S_1,S_2,\ldots,S_k\}$, where $S_j$ ($1\leq j \leq k$) represents the subspace covering all the video keyframes.
Given a query image $q\in \mathbb{R}^{d\times 1}$, the main objective of the query-by-image video task can be formulated as below:
\begin{equation}
{S^*} = \arg \mathop {\min }\limits_{S_j \in \mathcal{S}} d(q,S_j),
\end{equation}
where $d(\cdot,\cdot)$ is certain distance measurement of two data points. As shown, the major objective of query-by-image video retrieval is to find the subspace $S^*$ whose distance from query point $q$ is the shortest.

\subsection{Geometry-Preserving Distance}
The QBIVR is essentially a point-to-subspace search problem in which the query is represented as a point, and the database comprises subspaces.
Recall that existing solutions to the above problem either aggregate or project all the frames into a single datum compatible with the given query, which may cause serious information loss, such as spatial arrangement and/or temporal order. To compensate such drawbacks, we propose to measure the image-to-video relationship by the distance between the query and its corresponding projection on the subspace plane. In this way, the geometric property and structural information of subspace can be fully preserved. We denote the new image-to-video distance as
\begin{equation}
d(q,S) = \mathop {\min }\limits_{v \in S} \left\| {q - v} \right\|_2^2,
\label{eq:dqS}
\end{equation}
where $\|\cdot\|_2$ is the $\ell_2$ norm. It is easy to see that the nearest point is the orthographic projection of $q$ on $S$,
which is calculated as follows:
\begin{equation}
v^*=p(q)=\tilde{S}q,
\label{eq:sq}
\end{equation}
where $p(q)$ is the orthographic projection of $q$ on $S$, $\tilde{S}={S}({S}^{\top}{S})^{-1}{S}^{\top}$. Note that $\tilde{S}\in \mathbb{R}^{d\times d}$ can be computed offline to increase efficiency. Substituting Eq. (\ref{eq:sq}) into Eq. (\ref{eq:dqS}), we obtain the distance of point-to-subspace:
\begin{equation}
d(q,S)=  \| q-p(q)  \|_2  = \| (I_d-\tilde{S})q \|_2.
\end{equation}
where $I_d$ is an identity matrix of size $d\times d$. Denoting $D=I_d-\tilde{S}$, given that $\tilde{S}^{\top}\tilde{S}=\tilde{S}$ we have
\begin{equation}
D^{\top}D=(I_d-\tilde{S})^{\top}(I_d-\tilde{S})=I_d-\tilde{S}=D.
\end{equation}
Therefore, we can obtain the further conclusion
\begin{equation}
\begin{aligned}
d^{2}(q,S)&= \| Dq \|_2^{2}=Tr(D^{\top}Dqq^{\top})\\
&=Tr((I_d-\tilde{S})qq^{\top})=q^{\top}q-Tr(\tilde{S}qq^{\top}),
\end{aligned}
\label{eq:d2}
\end{equation}
where $Tr(\cdot)$ is the trace of a matrix. Based on Eq. (\ref{eq:d2}), our objective is equivalent to the following problem:
\begin{equation}
\begin{aligned}
S^*=&\arg\underset{S\in \mathcal{S}}{\max}~Tr(\tilde{S}qq^{\top})\\
\end{aligned}
\label{eq:S*}
\end{equation}
Noting that $Tr(\tilde{S}qq^{\top})=q^{\top}\tilde{S}q=\langle p(q),q \rangle$, where $\langle \cdot,\cdot \rangle$ denotes the inner product of two vectors. Then Eq. (\ref{eq:S*}) can be seen as a Maximum Inner Product Search (MIPS) problem w.r.t. the query $q$ and its orthographic projection $p(q)$ in the subspace. However, all the $p(q)$ in the dataset have to be preprocessed every time a new query is provided, which apparently increases the computational cost to solve. To bypass this issue,  based on the linear algebra manipulation $Tr(\tilde{S}qq^{\top})=\mathbf{vec}(\tilde{S})^{\top}\mathbf{vec}(qq^{\top})$, we rewrite the problem (\ref{eq:S*}) as in (\ref{eq:MIPS}):

\begin{equation}
S^*=\arg\underset{S\in \mathcal{S}}{\max}~\langle{\mathbf{vec}}(\tilde{S}),{\mathbf{vec}}(qq^{\top})\rangle,
\label{eq:MIPS}
\end{equation}
where $\mathbf{vec}(\cdot)$ is the function of transforming a matrix of size $d\times d$ to a column vector of size $d^2\times 1$ by performing column-wise concatenation of the matrix. In this way, we obtain an equivalent MIPS problem w.r.t. $\mathbf{vec}(\tilde{S})$ and $\mathbf{vec}(qq^{\top})$ from the original QBIVR problem.

Considering the unaffordable computation of $qq^{\top}$ and ${\mathbf{vec}}(\tilde{S})^{\top}{\mathbf{vec}}(qq^{\top})$ when $d$ is large, i.e., $O(d^2k)$, we employ hashing approaches to binarize image query and video data. Different properties of query images and videos in the database are unnegligible factors for accurate binary codes. Therefore, we learn asymmetric hash functions for image query $\mathbf{vec}(qq^{\top})$ and video data $\mathbf{vec}(\tilde{S})$ respectively, then the MIPS problem is reformulated as
\begin{equation}
S^*=\arg\underset{S\in \mathcal{S}}{\max}~\langle f(\mathbf{vec}(\tilde{S})),g(\mathbf{vec}(qq^{\top}))\rangle,
\label{eq:MIPS2}
\end{equation}
where $f\!:\!\mathbb{R}^{d^2\times 1}\!\rightarrow\! \{-1,1\}^{r\times 1}$ and $g\!:\!\mathbb{R}^{d^2\times 1}\!\rightarrow \!\{-1,1\}^{r\times 1}$ are hash functions for videos and images, respectively.

\subsection{Inner-product Binary Coding}
We first present the Inner-product Binary Coding approach. To facilitate the asymmetric learning of hash functions, we first construct video data $V$ and image data $U$ for training:
\begin{equation}\nonumber
\left\{
\begin{aligned}
V&\!=\![\mathbf{vec}(\tilde{S}_{1}),\mathbf{vec}(\tilde{S}_{2}),\ldots,\mathbf{vec}(\tilde{S}_{k})]\!\in\! \mathbb{R}^{d^2\times k}, \\ U&\!=\![\mathbf{vec}({x_1x_1^{\top}}),\mathbf{vec}({x_2x_2^{\top}}),\ldots,\mathbf{vec}({x_nx_n^{\top}})]\!\in\! \mathbb{R}^{d^2\times n},
\end{aligned}
\right.
\end{equation}
where $\{x_i\}|_{i=1}^n$ are $n$ images randomly sampled from video frames for training. Let $A$ be the correlation matrix of $U$ and $V$. We choose to use the inner product to represent the similarity, i.e., $A=U^{\top}V$. Following~\cite{AIBC15}, we now consider the following optimization problem:
\begin{equation}
\underset{f,g}{\min}\left \| g(U)^{\top}f(V)-A \right \|_F^{2},
\end{equation}
where $f(V)=[f(\mathbf{vec}(\tilde{S}_{1})),\ldots,f(\mathbf{vec}(\tilde{S}_{k}))]$, $g(U)=[g(\mathbf{vec}({x_1x_1^{\top}})),\ldots,g(\mathbf{vec}({x_nx_n^{\top}}))]$, and $\|\cdot\|_F$ is the Frobenius norm.
For simplicity, we choose to learn linear hash functions, i.e., $f(x)=\mathbf{sgn}(P^{\top}x)$ and $g(z)=\mathbf{sgn}(Q^{\top}z)$, where $P\in \mathbb{R}^{d^2\times l}$ and $Q\in \mathbb{R}^{d^2\times l}$ are the two mapping variables for binarizing videos and images, respectively.

In practice, to further speed up the optimization, we deliberately discard the quadratic term $\| g(U)^{\top}f(V)\|_F^{2}$, in view of the quadratic term with no help in leveraging the ground-truth similarity. In fact, the term $\| g(U)^{\top}f(V)\|_F^{2}$ can be treated a regularization in the magnitude of the learned inner product. Hence, we arrive at the new objective:
\begin{equation}
\underset{f,g}{\max}~Tr(g(U)Af(V)^{\top}),
\end{equation}
which can be optimized by alternatingly updating $g$ and $f$. In particular, when learning $g$ with $f$ fixed, we have
\begin{equation}
\label{eq:P}
\underset{P}{\max}~Tr(\mathbf{sgn}(P^{\top}U)Af(V)^{\top}).
\end{equation}
When updating $f$ with $g$ fixed, we arrive at
\begin{equation}
\label{eq:Q}
\underset{Q}{\max}~Tr(\mathbf{sgn}(Q^{\top}V)A^{\top}g(U)^{\top}).
\end{equation}
Both of the above sub-problems are of the same form. Next, we will show how to solve (\ref{eq:P}) and the sub-problem (\ref{eq:Q}) can be solved in the same way.

It is non-trivial to optimize the sub-problem (\ref{eq:P}) due to the existence of the sign function $\mathbf{sgn}(\cdot)$. To bypass the obstacle, we introduce an auxiliary variable $B\in \{-1,1\}^{r\times n}$ to approximate $\mathbf{sgn}(P^{\top}U)$, and thus we have
\begin{equation}
\begin{aligned}
\underset{B,P}{\max}~&Tr(BAf(V)^{\top})-\lambda \left \| B-P^{\top}U \right \|_F^{2},\\
\textrm{s.t.}~&B\in \{ -1,1 \}^{r\times n},
\end{aligned}
\end{equation}
where $\lambda>0$ is a balance parameter. Setting the derivative of the above objective w.r.t. $P$ to zero, we have
\begin{equation}\label{eq:P2}
P=(UU^{\top})^{-1}UB^{\top}.
\end{equation}
Fixing $P$, then we can update $B$ with
\begin{equation}\label{eq:B}
B=\mathbf{sgn}(f(V)A^{\top}+2\lambda P^{\top}A).
\end{equation}
The above analytical solution of $B$ significantly reduces the training cost, similarly making the algorithm easily performed on large-scale databases.

\subsection{Bilinear Binary Coding}
Note that in IBC, hashing the vectored images and videos data $\mathbf{vec}(\tilde{S})$ and $\mathbf{vec}(xx^{\top}\!)$ with a full projection matrix may cause high computational cost. In this part, we propose a Bilinear Binary Coding (BBC) approach to further accelerate the efficiency of QBIVR task.

We first present a bilinear rotation to maintain matrix structure instead of a single large projection matrix, denoted as $H(X)=\mathbf{sgn}(R_{1}^{{\top}}XR_{2})$, which is remarkable successful in lowering running time and storage for code generation. It also has been proved 		by~\cite{GongKRL13} that a bilinear rotation to $X\in \mathbb{R}^{d_{1}\times d_{2}}$ is equivalent to a $d_{1}d_{2}\times d_{1}d_{2} $ rotation to $\mathbf{vec}(X)$, denoted as ${R}=R_{2}\bigotimes R_{1}$, where $\bigotimes $ is the Kronecker product~\cite{0012754}. Now, we can equivalently learn asymmetric hash functions for images and videos as follows:
\begin{equation}
\small{
\left\{
\begin{aligned}
&f(\mathbf{vec}(\tilde{S}))=\mathbf{sgn}({P}^{{\top}}\mathbf{vec}(\tilde{S}))=\mathbf{sgn}(\mathbf{vec}(P_{1}^{\top}\tilde{S}P_{2})),\\
&g(\mathbf{vec}(xx^{\top}\!))\!=\!\mathbf{sgn}({Q}^{\top}\mathbf{vec}(xx^{\top}\!))\!=\!\mathbf{sgn}(\mathbf{vec}(Q_{1}^{\top}xx^{\top}\!Q_{2})).
\end{aligned}
\right.}
\end{equation}
where $P_{1},Q_{1}\in \mathbb{R}^{d\times c_{1}}$, $P_{2},Q_{2}\in \mathbb{R}^{d\times c_{2}}$,  $c_{1},c_{2}<d$. Then, we can generate binary codes for vectored images and videos with code length $c=c_{1}\times c_{2}$ for performing an efficient retrieval.

Following \cite{TPAMI.2012.193}, a feasible objective is to learn a bilinear rotation which minimizes the angle between
$\mathbf{vec}(R_{1}^{\top}XR_{2}) $ and its binary encoding $B=\mathbf{sgn}(\mathbf{vec}(R_{1}^{\top}XR_{2}))$. We preprocess the video dataset $V$ and image dataset $U$ to be zero-centered and have unit norm, then our goal is to maximize the following objective:
\begin{equation}
\sum_{i=1}^{n}\cos\theta^{(U)}_{i}+\sum_{j=1}^{k} \cos\theta^{(V)}_{j}+ \mu \sum_{i=1}^{n}\sum_{j=1}^{k}\delta _{i,j}\cdot\cos\omega_{i,j},
\end{equation}
where $\theta^{(J)}_{m}$ is the angle of the $m$-th rotated image/video and its binary code. $J\in \{U,V\}$. $\omega_{i,j}$ is the angle between the binary codes of $B_{i}^{(U)}$ and $B_{j}^{(V)}$, where $B_{i}^{(U)}=\mathbf{sgn}(Q_{1}^{\top}U_{i}Q_{2})$ and $B_{j}^{(V)}=\mathbf{sgn}(P_{1}^{\top}V_{j}P_{2})$ are the binary codes of the $i$-th image and $j$-th video, respectively. $\delta _{i,j}$ preserves the similarity property of images and videos in the different or same category:
\begin{equation}\nonumber
\delta _{i,j}=\left\{
\begin{array}{ll}
1, & U_{i},V_{j} \in \textrm{the same catagary},  \\
0, & \textrm{otherwise}.
\end{array}\right.
\end{equation}
For images, $\cos \theta^{(U)}_{i}$ is expressed as:
\begin{equation}
\cos \theta^{(U)}_{i}\!=\!\frac{\mathbf{vec}(\mathbf{sgn}(Q_{1}^{\top}U_{i}Q_{2}))^{\top}}{\sqrt{c}}\frac {\mathbf{vec}(Q_{1}^{\top}U_{i}Q_{2})}{\left \| \mathbf{vec}(Q_{1}^{\top}U_{i}Q_{2}) \right \|_{2}}.
\end{equation}
To simplify the subsequent optimization, we follow~\cite{GongKVL12} to relax the above objective function by ignoring $\left \| \mathbf{vec}(Q_{1}^{\top}U_{i}Q_{2}) \right \|_{2}$, and arrive at:
\begin{equation}
\begin{aligned}
&\sum\nolimits_{i}\cos\theta^{(U)} _{i}\\
=&\sum\nolimits_{i}(\frac{\mathbf{vec}(\mathbf{sgn}(Q_{1}^{\top}U_{i}Q_{2}))^{\top}}{\sqrt{c}}\mathbf{vec}(Q_{1}^{\top}U_{i}Q_{2}))\\
=&\frac{1}{\sqrt{c}}\sum\nolimits_{i}(\mathbf{vec}(B^{(U)}_{i})^{\top}\mathbf{vec}(Q_{1}^{\top}U_{i}Q_{2}))\\
=&\frac{1}{\sqrt{c}}\sum\nolimits_{i}Tr(B^{(U)}_{i}Q_{2}^{\top}U_{i}^{\top}Q_{1}).
\end{aligned}
\end{equation}
Similarly, we can derive the objectives of video angles and image-to-video angles:
\begin{equation}
\left\{
\begin{aligned}
&\sum\nolimits_{j}\cos\theta^{(V)}_{j}=\frac{1}{\sqrt{c}}\sum\nolimits_{j} Tr(B_{j}^{(V)}P_{2}^{\top}V_{j}^{\top}P_{1}),\\
& \sum\nolimits_{i,j}\delta _{i,j}\cdot \cos\omega _{i,j}=\frac{1 }{c}\sum\nolimits_{i,j} \delta _{i,j}\cdot Tr(B_{j}^{(V)}(B_{i}^{(U)})^{\top}).
\end{aligned}
\right.
\end{equation}
Hence, the objective function is transformed to
\begin{equation}
\begin{aligned}
  &\sum\nolimits_{i}Tr(B^{(U)}_{i}Q_{2}^{\top}U_{i}^{\top}Q_{1})+\sum\nolimits_{j} Tr(B_{j}^{(V)}P_{2}^{\top}V_{j}^{\top}P_{1})\\
  &+\frac{\mu }{\sqrt{c}}\sum\nolimits_{i,j} \delta _{i,j}\cdot Tr(B_{j}^{(V)}(B_{i}^{(U)})^{\top})
\end{aligned}
\end{equation}
where $ B_{i}^{(U)},B_{j}^{(V)}\in \{ -1,1 \}^{d\times d}$. $Q_{1}^{\top}Q_{1}=I,Q_{2}^{\top}Q_{2}=I,P_{1}^{\top}P_{1}=I,P_{2}^{\top}P_{2}=I$.

For optimization, we use block coordinate ascent to alternatingly update $ \{ B_{j}^{(V)}  \}_{j=1}^k$,$ \{B_{i}^{(U)} \}_{i=1}^n$,$Q_{1},Q_{2},$$P_{1},P_{2}$. The updating processes w.r.t. images and videos are symmetric. Hence we just describe the updates of variables of videos by fixing all the variables of images.

\noindent \textbf{Step 1:} Update $P_{1}$, with all other variables fixed. We have the following reduced problem:
\begin{equation}
\max_{P_1^{\top}P_1=I}~Tr(D_{1}P_{1}),
\label{eq:S1}
\end{equation}
where $D_{1}=\sum_{j=1}^{k}(B_{j}^{(V)}P_{2}^{\top}V_{j}^{\top})$. We can solve the above optimization problem using polar decomposition: \begin{equation}
  P_{1}=Y_{1}Z_{1}^{\top},
\end{equation}
where $Z_{1}$ and $Y_1$ are the left-singular vectors and the top $c_1$ right-singular vectors of $D_{1}$, respectively, by performing SVD.

\noindent \textbf{Step 2:} Update $P_{2}$, with all the others fixed, we have
\begin{equation}
\max_{P_2^{\top}P_2=I}~Tr(P^{\top}_{2}D_{2}),
\label{eq:S2}
\end{equation}
where $D_{2}=\sum_{j=1}^{k}(V_{j}^{\top}P_{1}B_{j}^{(V)})$. Similar to the previous step, the update for $P_{2}$ is $P_{2}=Z_{2}Y_{2}^{\top}$, where $Z_{2}$ and $Y_2$ are the top $c_2$ left-singular vectors and the right-singular vectors of $D_{2}$, respectively, by performing SVD.

\noindent \textbf{Step 3:} Update $B_{j}^{(V)}$, by fixing all the other variables, we obtain
\begin{equation}
\max_{B_{j}^{(V)}\in \{-1,1\}^{d\times d}}~Tr(B_{j}^{(V)}D_{3}),
\label{eq:S3}
\end{equation}
where $D_{3}\!=\!P_{2}^{\top}V_{j}^{\top}P_{1}\!+\!\frac{\mu}{\sqrt{c}} \sum\nolimits_{i}\delta _{i,j}(B_{i}^{(U)})^{\top}$. It can be easily seen that the solution to the above problem is as below:
\begin{equation}
  B_{j}^{(V)}=\mathbf{sgn}(D_{3}^{\top}).
\end{equation}
Then, we can similarly update $Q_1$, $Q_2$ and $\{B_{i}^{(U)} \}_{i=1}^n$.

Comparing to the time complexity of full rotation ,i.e, $O(d^{2})$, the asymmetric bilinear hashing learning of videos and images significantly reduces the training cost to $O(d_{1}^{2}+d_{2}^{2})$, where $d=d_{1}\times d_{2}$. We summarize the algorithm for optimizing the proposed Bilinear Binary Coding (BBC) approach in Algorithm \ref{alg:1}.
\begin{algorithm}[!ht]
	\begin{algorithmic}[1]
		\renewcommand{\algorithmicrequire}{\textbf{Input:}}
		\renewcommand{\algorithmicensure}{\textbf{Output:}}
		\REQUIRE Subspaces of videos $\{S_{j}\}|_{j=1}^k$ and images $\{x_{i}\}|_{i=1}^n$;
		\ENSURE Hash function $f$ and $g$;
		\STATE Compute $\tilde{S}_j={S}_j({S}_j^{\top}{S}_j)^{-1}{S}_j^{\top}$, $j=1,2,\ldots,k$;
		\STATE Construct video and image training data as below:
		
		\[
		\left\{
		\begin{aligned}
		V\!&=\!\left \{ \tilde{S}_{1},\tilde{S}_{2},\ldots,\tilde{S}_{k}\right \},\\
		U\!&=\!\left \{x_1x_1^{\top},x_2x_2^{\top},\ldots,x_nx_n^{\top}\right \};
		\end{aligned}
		\right.
		\]
		
		\STATE Randomly initialize $ \{ B_{j}^{(V)}  \}_{j=1}^k$,$ \{B_{i}^{(U)} \}_{i=1}^n$,$Q_{1},Q_{2},P_{2}$;
\vspace{-0.5cm}
		 \REPEAT{
			\STATE Update $P_{1}$ by solving the problem (\ref{eq:S1});
			\STATE Update $P_{2}$ by solving the problem (\ref{eq:S2});
			\STATE Sequentially update $\{ B_{j}^{(V)}  \}_{j=1}^k$ by solving the problem (\ref{eq:S3});
			\STATE Update $Q_{1}$ according to the problem (\ref{eq:S1});
			\STATE Update $Q_{2}$ according to the problem (\ref{eq:S2});
			\STATE Sequentially update $\{B_{i}^{(U)} \}_{i=1}^n$ according to the problem (\ref{eq:S3});
		}
		\UNTIL{there is no change to all the variables;}
		\RETURN{$ \{ B_{j}^{(V)}  \}_{j=1}^k$,$ \{B_{i}^{(U)} \}_{i=1}^n$,$Q_{1},Q_{2},P_1,P_{2}$.}
	\end{algorithmic}
	\caption{Optimization of Bilinear Binary Coding.}
	\label{alg:1}
\end{algorithm}

\section{Experiments}
In this section, we evaluate our two proposed IBC, BBC approaches on four datasets for the query-by-image video retrieval (QBIVR) task.
\subsection{Data and Experimental Settings}
We used four video datasets, a face video dataset BBT1 (Big Bang Theory1)~\cite{LiWHSC15}, UQE50 (UQ Event dataset with 50 pre-defined events) dataset~\cite{Yu2016}, a micro object-based video dataset collected by ourselves from Vine\footnote{https://vine.co/}, and a wide range of objects and events dataset Fudan-Columbia Video Dataset (FCVID)\footnote{http://bigvid.fudan.edu.cn/FCVID/}. BBT1 dataset, a low-dimensional small dataset, has been proved by HER method in ~\cite{LiWHSC15}. We conduct experiments on this publicly-available video dataset to verify the effectiveness of our approaches. For the other datasets, we first adopted the FFmpeg\footnote{http://ffmpeg.org/} to sample the videos at the rate of $5$ frames per second as keyframes, and subsequently extracted the visual features of keyframes using fc7 layer ($4096$-d) of VGG Net model~\cite{XuYH15}. In view of the potential redundancy, the data in our experiment were further reduced to $1600$-d by PCA.

We compared our IBC and BBC approaches against several state-of-the-art unsupervised hashing methods for large-scale video retrieval, including ALSH~\cite{Shrivastava014}, IMH~\cite{Song2013Inter}, SH~\cite{SH08}, ITQ~\cite{TPAMI.2012.193}. In our IBC approach, each column of the original inner product matrix $A$ is binarized, where the top $m$ largest elements are set to $1$ and the rest ones to $0$. $m$ is set to $10,000$ for the Vine dataset, $1,500$ for UQE50 and BBT1 datasets, and $18,000 $ for FCIVD dataset. Notably, the balance parameter $\lambda$ is empirically set to $100$ and the number of local iterations $t$ is set to $2$. In our BBC approach, we firstly initiate the bilinear rotation parameters $W_{1},W_{2},P_{1},P_{2}$ randomly and then learn two asymmetric hash functions respectively.
The number of local iterations $iter$ is set to $10$ in light of the excellent converging property of our devised approach. The parameters of the rest compared approaches are set as suggested above. In the experiment, the code length is tested in the range of $\{16, 32, 64, 96,128\}$.

The evaluation metrics are chosen as Hamming ranking including mean of average precision (mAP) and mean precision of the top $500$ retrieved neighbors (Precision@500).

\subsection{BBT1: video retrieval with face images}
 The Big Bang Theory (BBT)~\cite{LiWHSC15} is a sitcom (20 minutes an episode) which includes many full-view shots of multiple characters at a time. It take places mostly indoors and mainly focuses on $5~8$ characters. BBT1 consists of 3,341 face videos of the first 6 episodes from season 1 of BBT which are represented by block Discrete Cosine Transformation (DCT) feature as used in~\cite{BaumlTS13}, which forms a $240 \times 240$ covariance video representation.
\subsubsection{Compared to other state-of-the-art methods}
To compare with HER method~\cite{LiWHSC15}, an effective heterogeneous spaces hashing method, we tested our approaches on the BBT1 which has been verified successfully on HER method.
Following~\cite{LiWHSC15}, for each database, we randomly extracted 300 image-video pairs (both elements of the pair come from the same subject) for training and 100 images from the rest as query for the retrieval task. The results are shown in Table\ref{tab:HER}. Our proposed approaches not only perform better than HER method on mAP, but also overcome the limitation of HER method in low dimension and the subsequent experiments show that our approaches can be applied to high-dimensional large datasets fittingly.
\begin{table}[!htb] 
	\centering 
	\begin{tabular}{c|cccc}
		\hline
		\multirow{2}{*}{\textbf{Method}}  &\multicolumn{4}{c}{\textbf{mAP}} \\ \cline{2-5}
		& \textbf{16-bit}& \textbf{32-bit}& \textbf{64-bit}& \textbf{96-bit} \\  \hline\hline
		HER & 0.5049 &0.5227 &0.5490 &0.5531\\ 
		IBC & 0.5152 &0.5369 &0.5561 &0.5638 \\
		BBC & 0.5080&0.5401 &0.5643 &0.5711\\ 
		\bottomrule
	\end{tabular}
	\caption{Comparison of HER method and IBC, BBC approaches on BBT1 dataset with $16,32,64,96$-bit.} 
	\label{tab:HER} 
\end{table}
\subsection{Vine: video retrieval with object images}

Vine is a micro video sharing platform, where users can only share videos which are no more than six seconds by mobile devices. We collected a micro object-based video dataset from Vine comprising $498,000$ micro videos in $145$ categories, and randomly sampled 11,000 videos and 11,000 images from videos for training respectively and the rest 1,000 images as test.

\subsubsection{Compared to other state-of-the-art methods}
We report the compared results with some state-of-the-art methods, i.e., hashing methods and vector-based approach in terms of both hash lookup: mAP and Precision@500. The compared vector-based method, a temporal aggregation~\cite{TAVR15} employs the Scalable Compressed Fisher Vectors (SCFV) to reduce retrieval latency and memory requirements for achieving higher speed and maintaining good performance. However, the approach sacrifices useful information such as structure similarity when pursuing a higher speed. Moreover, though binarized fisher features that TAVR uses are more representative and effective than low-level image features, it still fails when competing with deep features, especially the ones after redundancy removing. The performance of the vector-based approach and the proposed approaches in 64-bit is clearly illustrated in Table \ref{tab:vector}.
\begin{table}[!htb] 
	\centering 
	\begin{tabular}{l c c} 
		\toprule 
		\textbf{Method} &\textbf{mAP}  &\textbf{Precision@500}  \\ 
		\midrule 
		TAVR & 0.3785 &0.3021 \\ 
		IBC & 0.4997&0.4990 \\
		BBC & \textbf{0.5045}&\textbf{0.5039} \\ 
		\bottomrule
	\end{tabular}
	\caption{Comparison of vector-based retrieval and IBC, BBC on Vine dataset with code length is set to $64$.} 
	\label{tab:vector} 
\end{table}

In the comparisons with hashing methods, we treat a query a false case if no point is returned when calculating precision. Ground truths are defined by the category information from the datasets. As Table \ref{tab:Vine} shows, the two proposed approaches outperform all the other state-of-the-art methods in terms of every metric at all code lengths, noticeably remaining much greater expression ability when encoding length is as large as 128 bits.

\begin{table*}[!htb]
	\centering
	\resizebox{\linewidth}{!}{
		\def\arraystretch{1.2}
		\begin{tabular}{c|ccccc|ccccc}
			\hline
			\multirow{2}{*}{\textbf{Method}}  &\multicolumn{5}{c|}{\textbf{mAP}}  &\multicolumn{5}{c}{\textbf{Precision@500}}  \\ \cline{2-11}
			& \textbf{16-bit}& \textbf{32-bit}& \textbf{64-bit}& \textbf{96-bit}& \textbf{128-bit}& \textbf{16-bit}&\textbf{32-bit}& \textbf{64-bit}& \textbf{96-bit}& \textbf{128-bit}\\  \hline\hline
			ALSH  	& 0.2965& 0.2983& 0.3006& 0.3067& 0.3122& 0.2942& 0.2976& 0.3014& 0.3142 &0.3224	\\ \hline
			ITQ 	& 0.3121& 0.3156& 0.3256& 0.3325& 0.3378& 0.2846& 0.3256& 0.3274& 0.3302 &0.3544	\\ \hline
			IMH 	& 0.3308& 0.3569& 0.3596& 0.3628& 0.3642& 0.3282& 0.3366& 0.3644& 0.3722 &0.3804	\\ \hline
			SGH 	& 0.3849& 0.4135& 0.4258& 0.4322& 0.4371& 0.4165& 0.4276& 0.4294& 0.4322 &0.4424	\\ \hline
			IBC 	& 0.4970& 0.4985& 0.4997& 0.5012& 0.5064& 0.4892& 0.4953& 0.4990& 0.5010 &0.5048	\\ \hline
			BBC 	&\textbf{0.4933}& \textbf{0.5012}& \textbf{0.5045}& \textbf{0.5082}& \textbf{0.5125}& \textbf{0.4842}&\textbf{0.4993}& \textbf{0.5039}& \textbf{0.5115}& \textbf{0.5202}	 \\ \hline
		\end{tabular}}
		\caption{Comparison of IBC, BBC and other hashing methods on Vine with code lengthes are set to $16$,$32$,$64$,$96$ and $128$.}
		\label{tab:Vine}
	\end{table*}


\subsection{UQE50: video retrieval with event images}
The video dataset UQE50 (UQ Event dataset with 50 pre-defined events) aims at event analysis tasks which was downloaded from YouTube\footnote{http://www.youtube.com/}~\cite{Yu2016}.
The dataset contains $3,462$ videos that belong to $50$ different event categories, and all the videos are from trending events happened in the last few years whose granularity is comparably larger than the existing video event datasets.
Compared with the Vine dataset of $100,000+$ object-based videos, UQE50 is a longer-time event-based video dataset in a smaller size. To verify the generality of our proposed approaches in different typies of video retrieval, we used UQE50 video dataset to compare the performance of our approaches with that of other state-of-the-art methods.
We randomly chose $1,800$ videos and images from videos as training samples respectively and the rest 200 images were used as test samples.

\subsubsection{Compared with other state-of-the-art methods}
\begin{figure*}[!htb]
	\centering
	\subfloat[mAP on UQE50 dataset]{
		\includegraphics[width=0.25\linewidth]{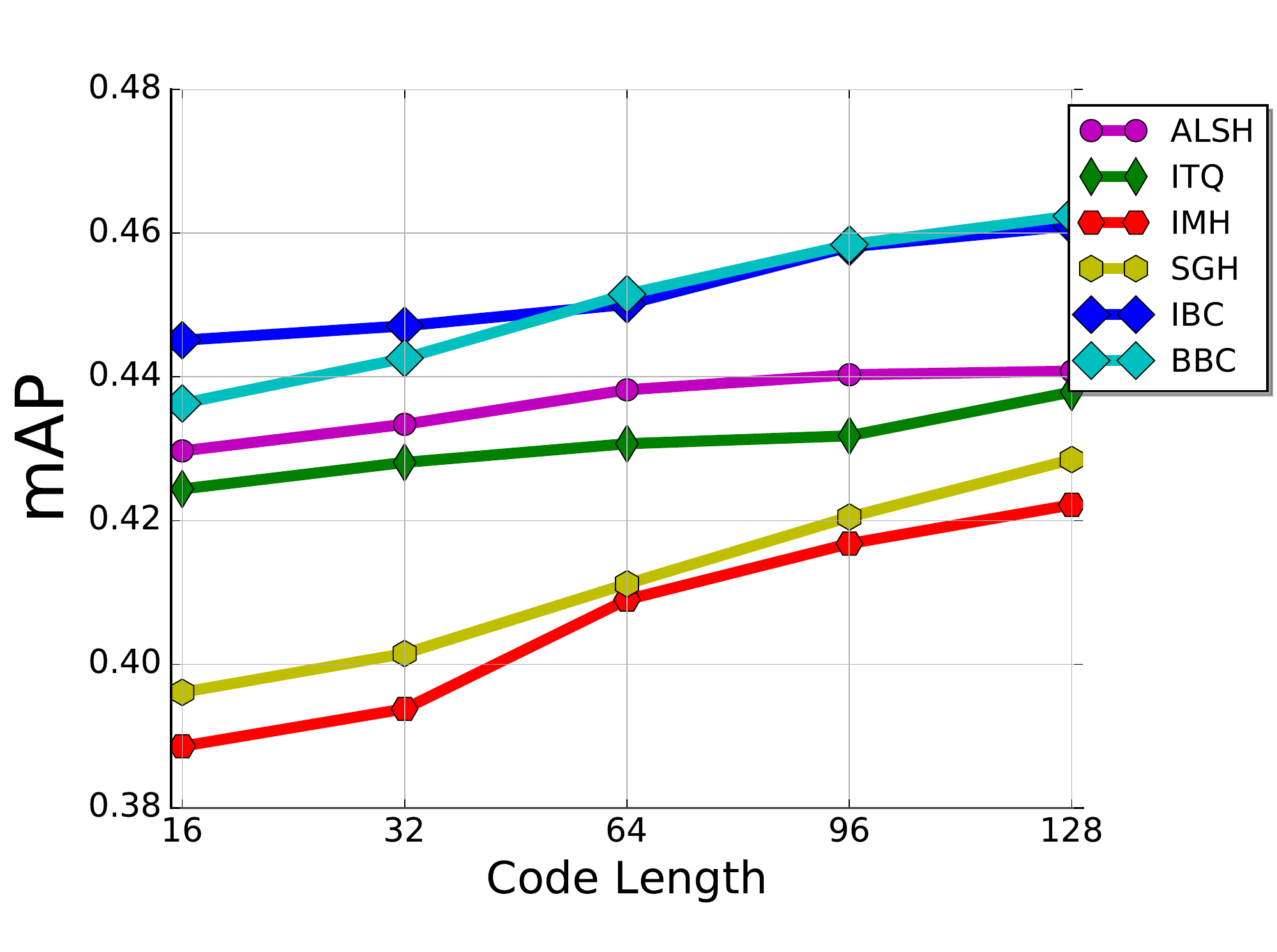}}
	\subfloat[Precision@500 on UQE50 dataset]{
		\includegraphics[width=0.25\linewidth]{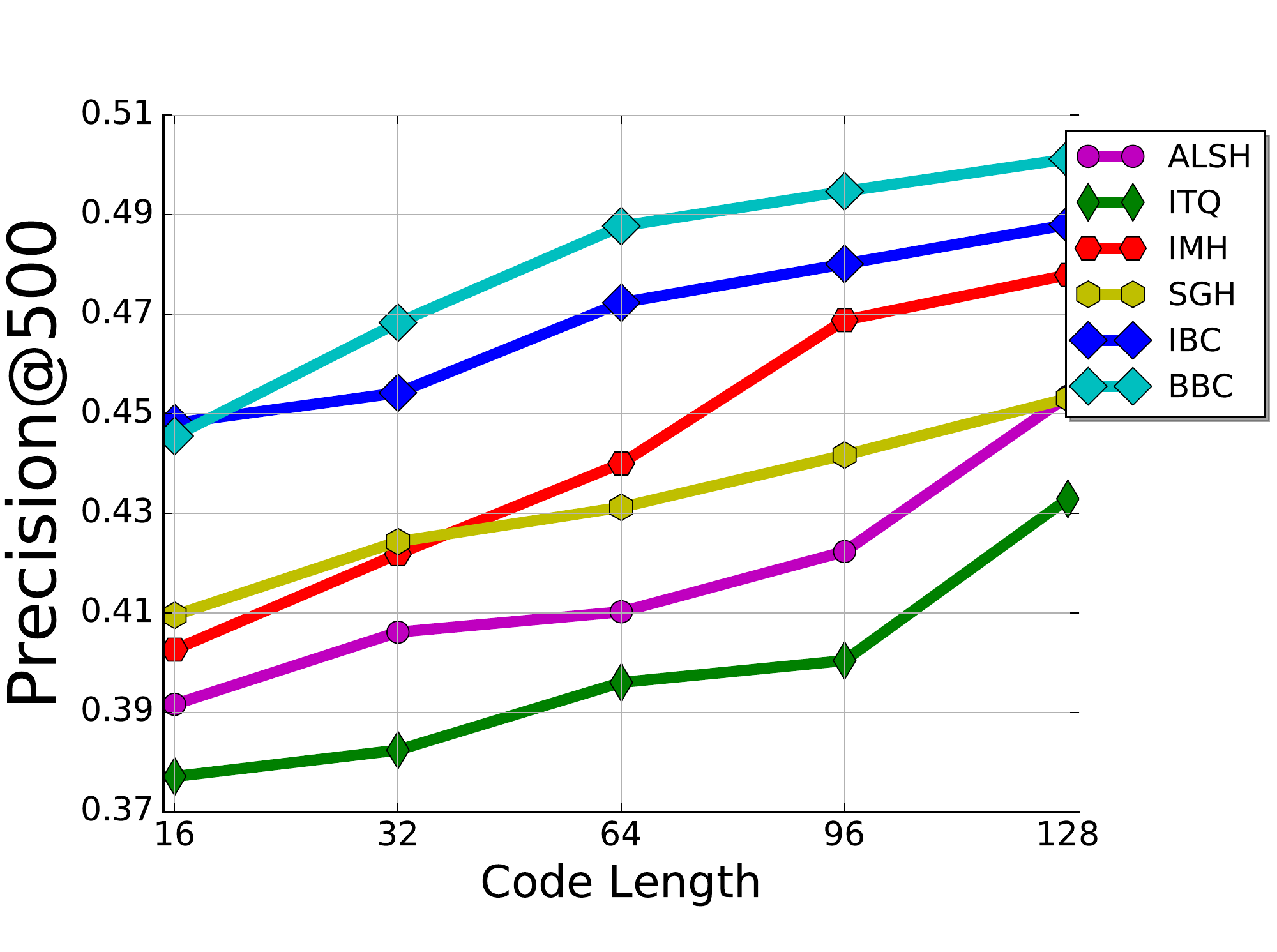}
	}
	\subfloat[mAP on FCVID dataset]{
		\includegraphics[width=0.25\linewidth]{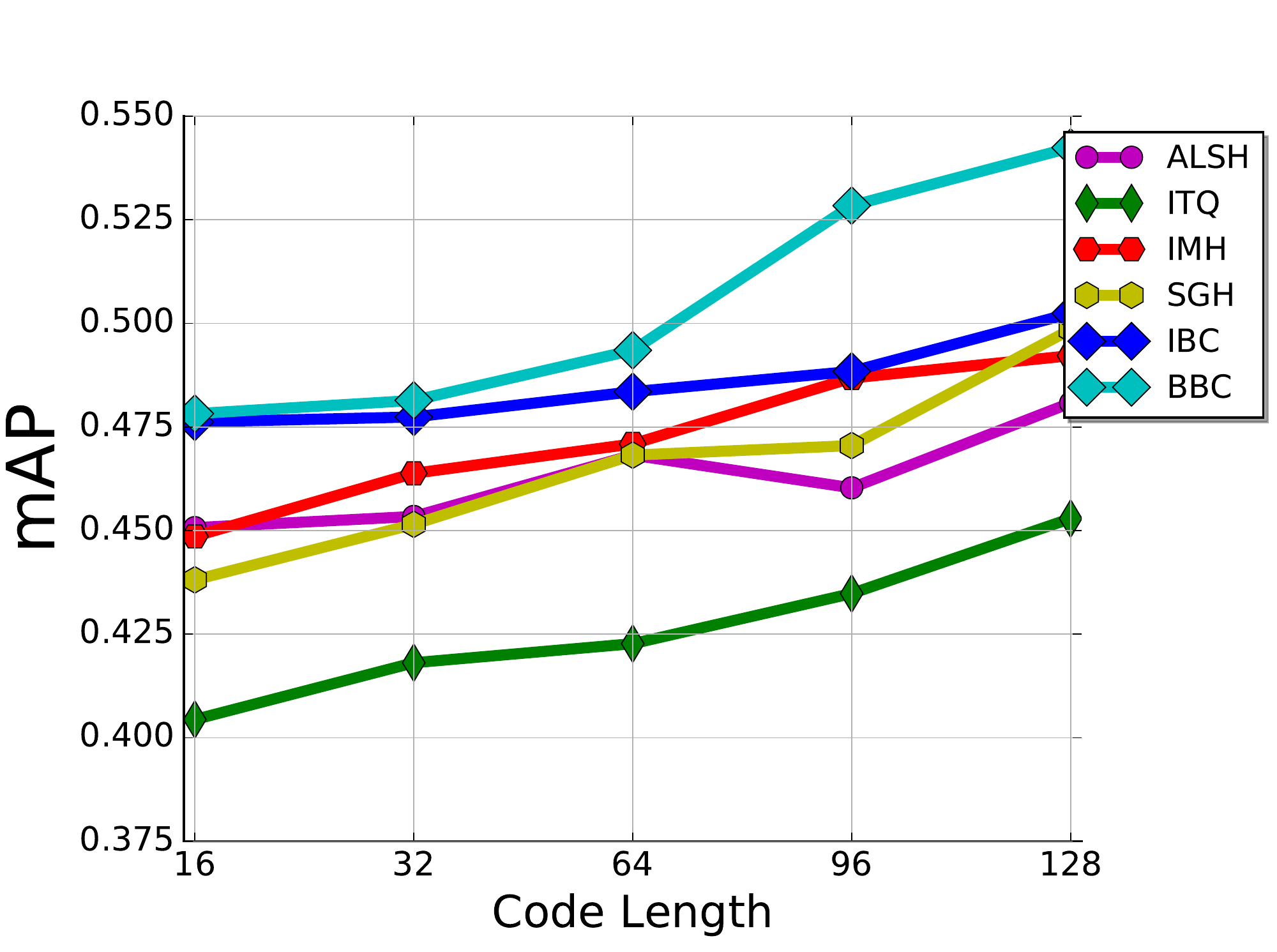}
	}
	\subfloat[Precision@500 on FCVID dataset]{
		\includegraphics[width=0.25\linewidth]{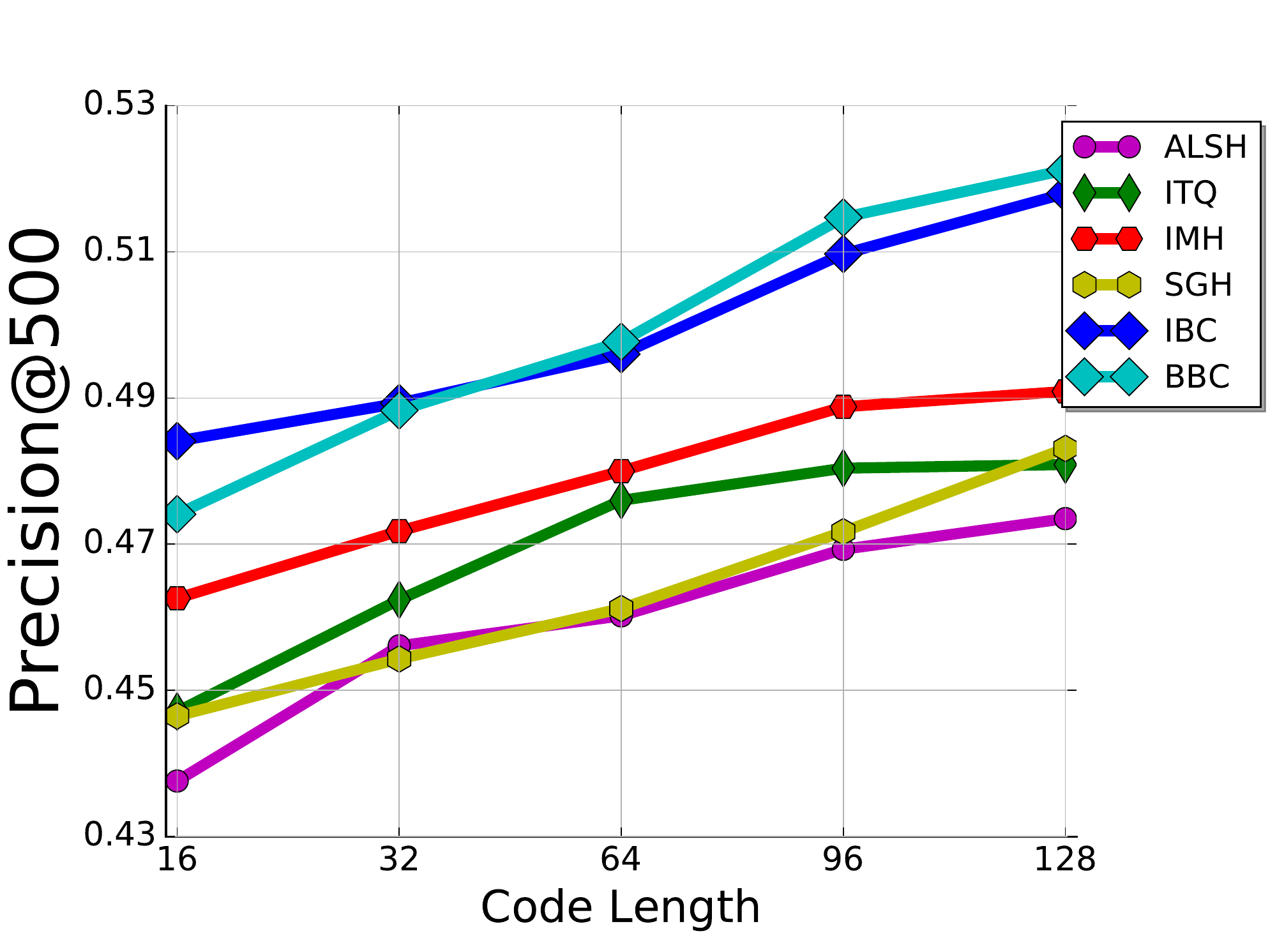}
	}
	
	\caption{Comparison of IBC, BBC and other hashing methods on FCVID dataset with different code lengths.}
	\label{fig:FCVID/UQE50}
\end{figure*}

To examine the practical efficacy of our proposed approaches, in this part, we conducted a similar experiment on UQE50 to evaluate its scalability compared with other state-of-the-art hashing methods as Vine experiment. Obviously, even for a smaller dataset of longer time-length, the performance (i.e., mAP, Precision@500) is also excellent as shown in Figure~\ref{fig:FCVID/UQE50}(a)(b). As the code length enlarges, all the aspects of the proposed approaches' performance steadily are better than the other ones.
	
	

\subsection{FCVID: video retrieval with event/object images}
The video dataset FCVID (Fudan-Columbia Video Dataset)\footnote{http://bigvid.fudan.edu.cn/FCVID/} is a web videos dataset containing $91,223$ Web videos annotated manually according to 239 categories. The categories in FCVID cover a wide range of topics like social events (e.g., ¡°tailgate party¡±), procedural events (e.g., ¡°making cake¡±), objects (e.g., ¡°panda¡±), scenes (e.g., ¡°beach¡±), etc. These categories were defined very carefully and organized in a hierarchy of 11 high-level groups. Specifically, the categories were conducted by user surveys and the organization structures on YouTube and Vimeo as references to identify.
In this section, we chose $40,000$ as our video dataset and randomly selected $22,000$ images and videos for training, then we tested another $5,000$ images in the video dataset.

\subsubsection{Compared with other state-of-the-art methods}
In this part, we conducted the same experiment as Vine and UQE50 to the FCVID dataset for studying the performance of event/object image retrieval. In view of the wide range of image typies and higher reliability of dataset, we can demonstrate the effect of our approaches clearly. As Figure\ref{fig:FCVID/UQE50} (c)(d) show, our approaches outperform both of mAP and Precision@500 in different code lengths than the other state-of-the-art hashing methods. Better performances on the four datasets in different types and sizes prove our validity of the proposed IBC and BBC approaches.
\begin{figure*}[!htb]
	\centering
	\subfloat[IBC on Vine]{
		\includegraphics[width=0.25\linewidth]{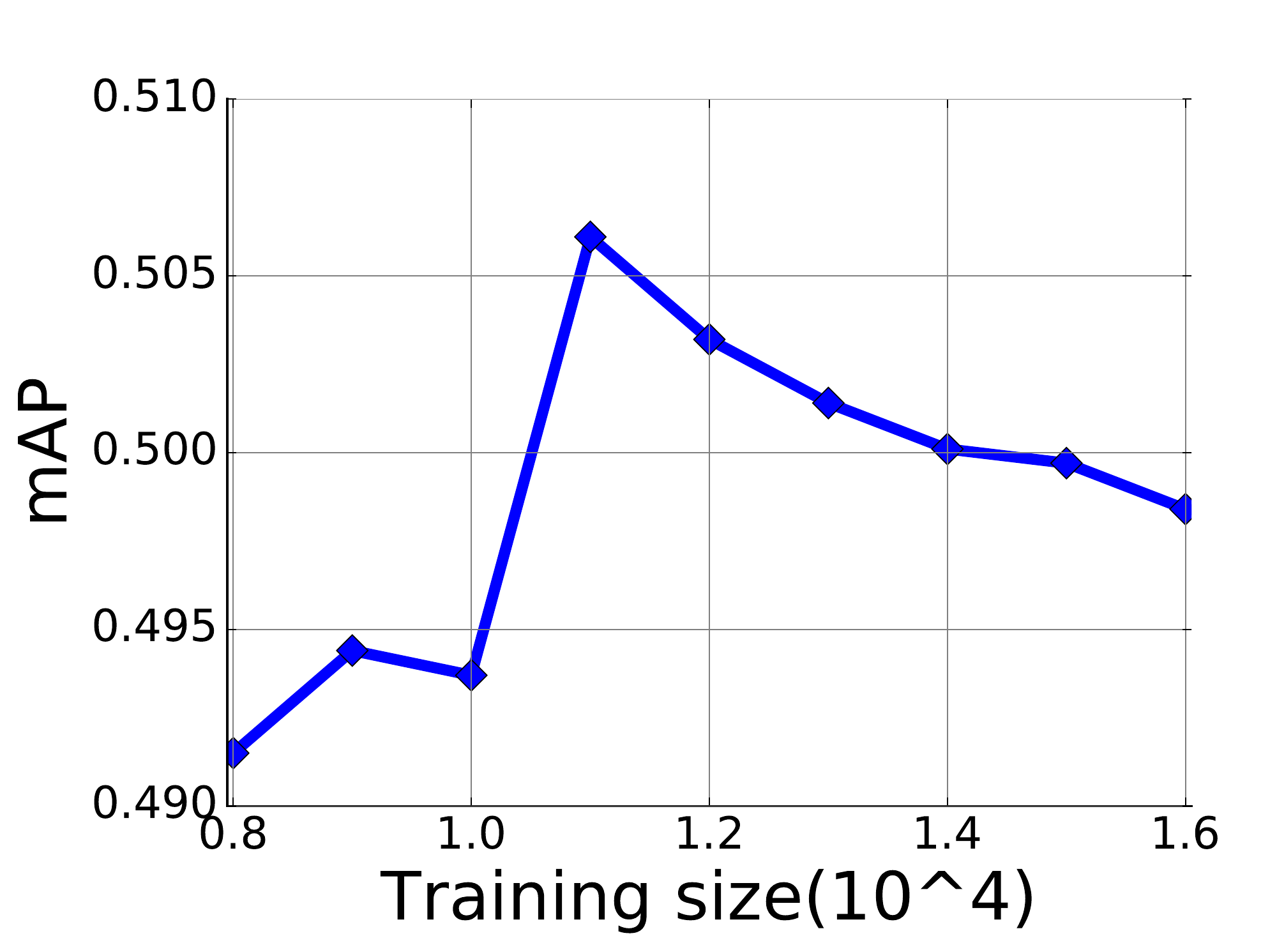}
	}
	\subfloat[BBC on UQE50]{
		\includegraphics[width=0.25\linewidth]{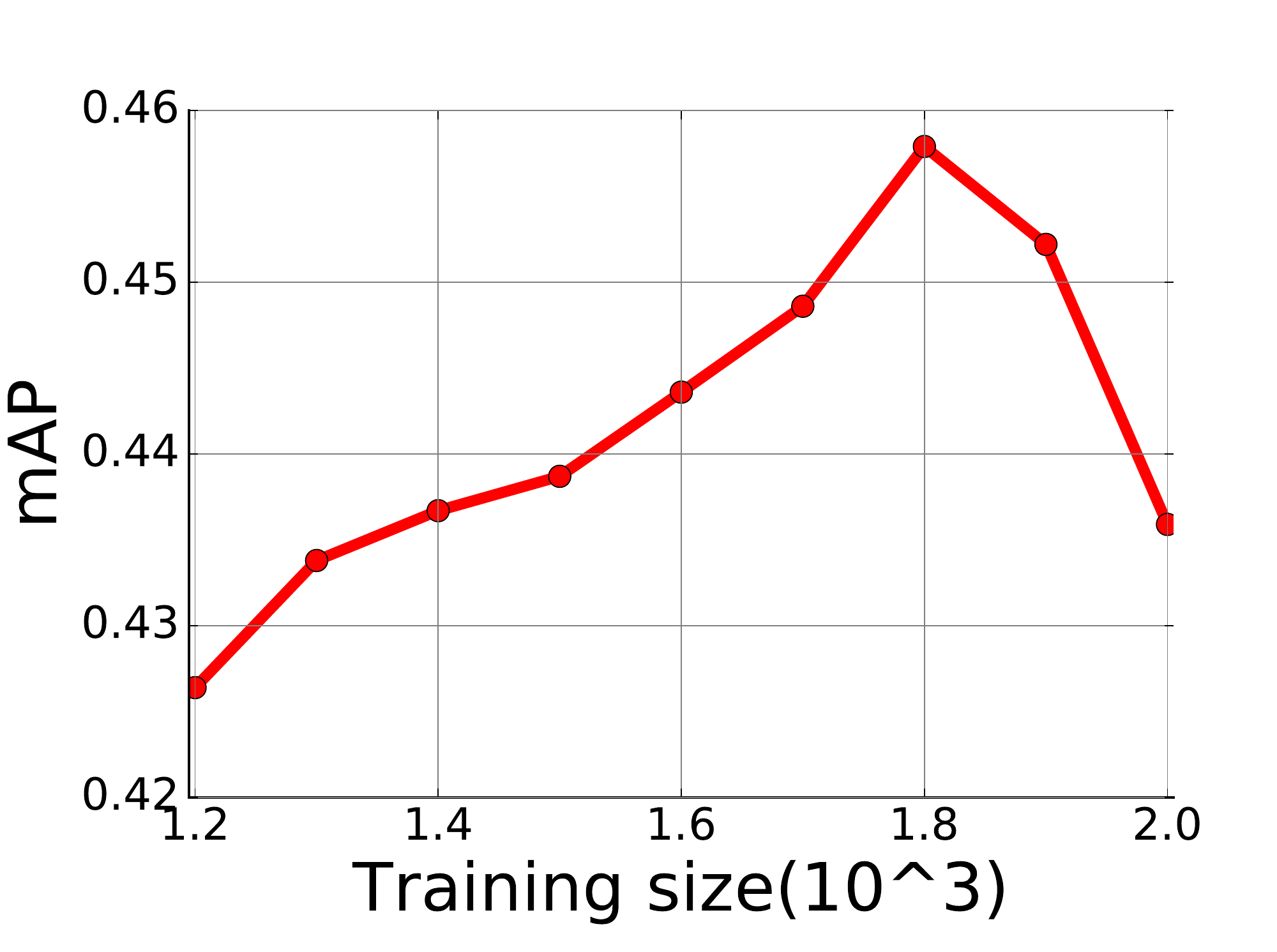}
	}
	\subfloat[IBC on Vine]{
		\includegraphics[width=0.25\linewidth]{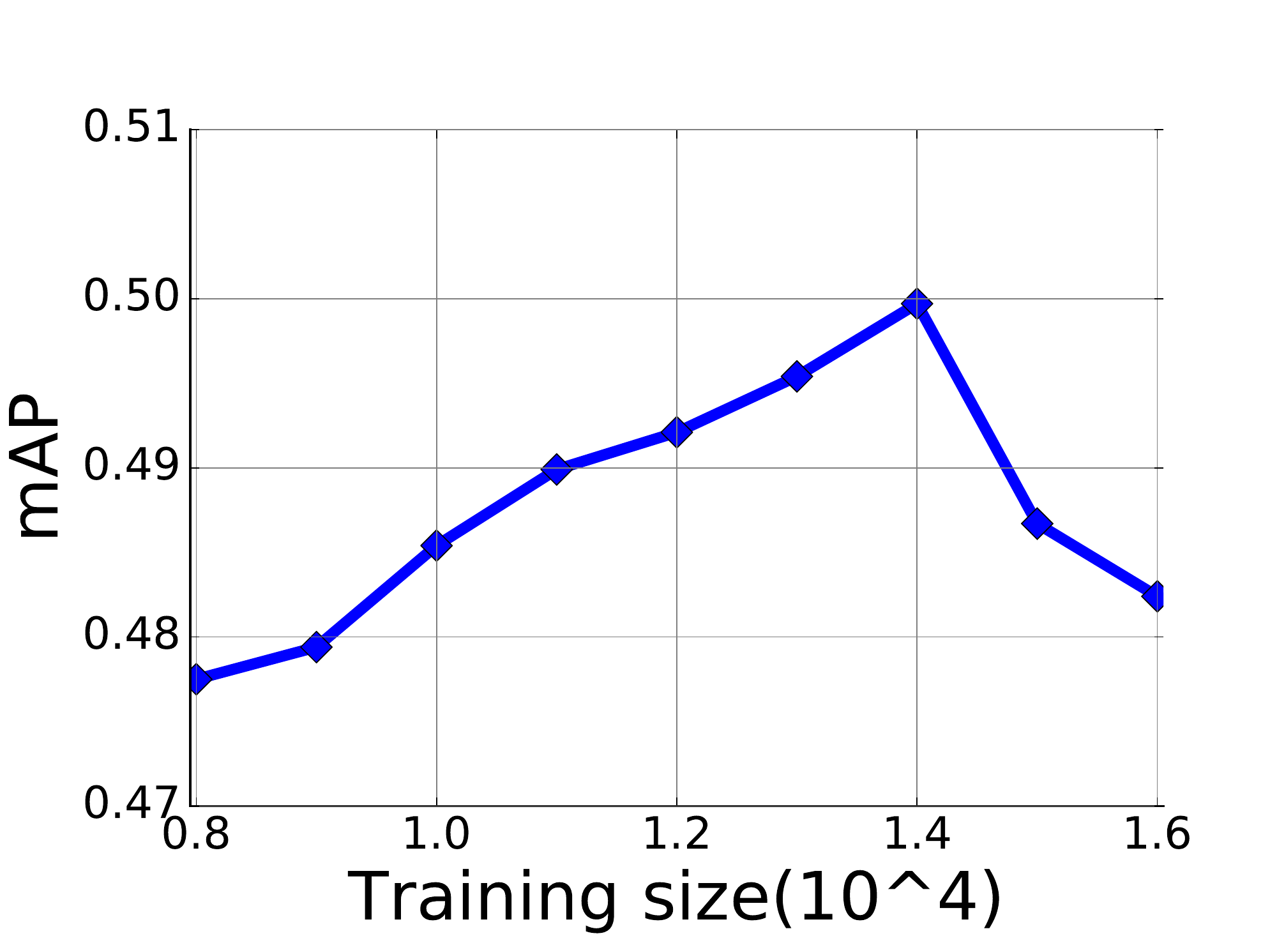}
	}
	\subfloat[BBC on UQE50]{
		\includegraphics[width=0.25\linewidth]{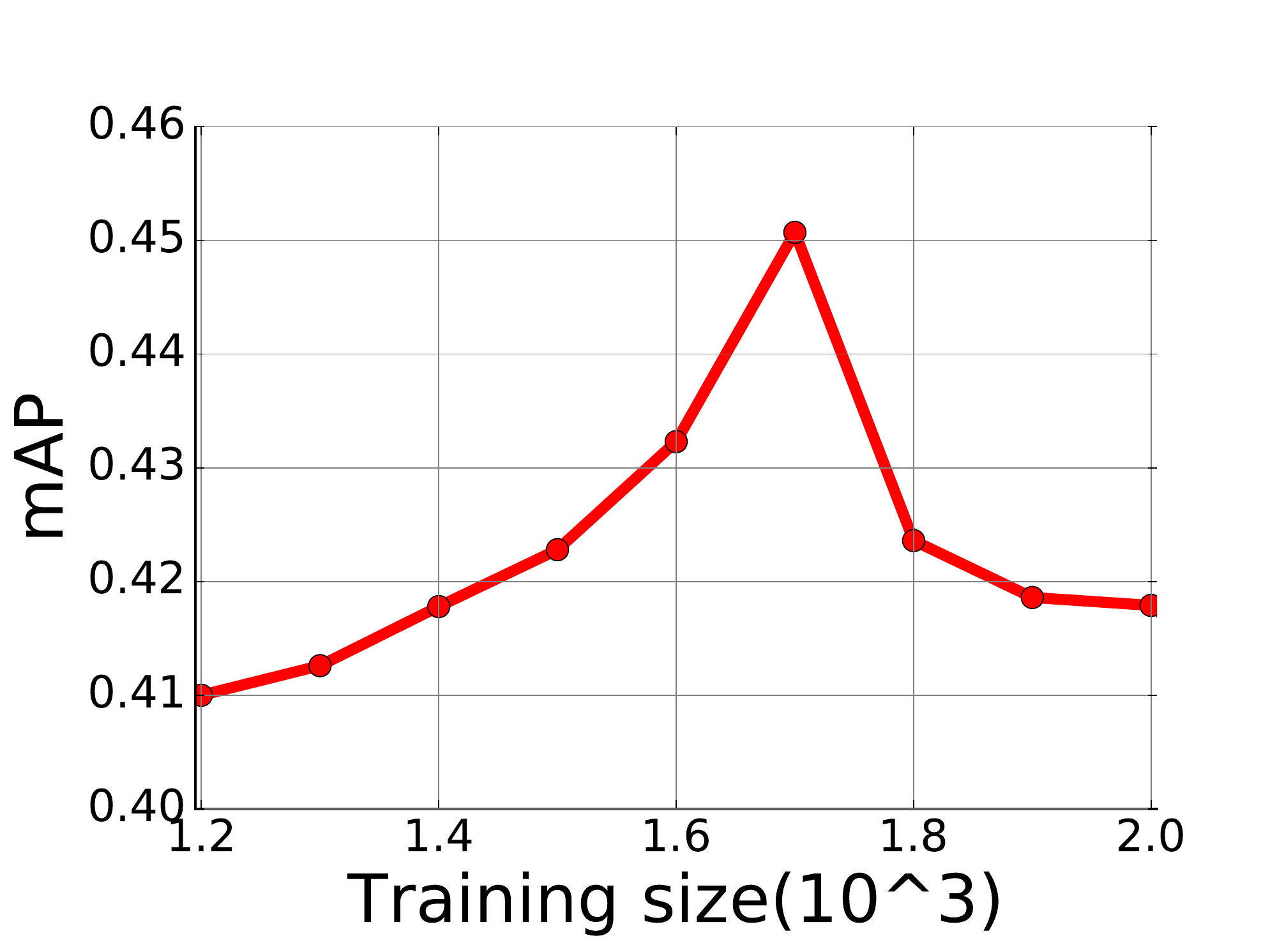}
	}
	\caption{Training size effects of IBC, BBC on mAP performance over Vine and UQE50 with 64 bits fixed.}
	\label{fig:UQsize}
\end{figure*}

\subsection{Effect of training size on UQE50 and Vine}
This part of experiment mainly studies on evaluating the effect of training size on the searching quality of our IBC and BBC approaches. We performed the experiments on object-based Vine and event-based UQE50 datasets and selected mean of average precision (mAP) as the comprehensive assessment index. We fixed the code as $64$-bit and varied the training size of Vine dataset from $8,000$ to $16,000$ with a regular interval of $1,000$ and UQE50 is tuned from $1,200$ to $2,000$ with a regular interval of $100$.
The subsequent consequences are shown in Figure~\ref{fig:UQsize}. As we can see, IBC and BBC approaches both have the suitable training size for the best performance, even though add more training data, the two approaches do not gain noticeable performance boost to some extent. The ideal training size of Vine is $14,000$ for IBC approach and $11,000$ for BBC approach. And for UQE50 dataset, the performance ultimately is optimal with the training size of $1,700$ for IBC approach and $1,800$ for BBC approach.
This section also gives us guidance for chosing the suitable training size.

\section{Conclusion}
In this paper, we developed Binary Subspace Coding (BSC) framework which includes two different approaches for query-by-image video retrieval. Different from traditional video retrieval methods, we focused on subspace-based video representation and discovered a common Hamming space for both images and videos, to enable an efficient retrieval. Our proposed similarity-preserving measurement can preserve  geometric structure properties of videos than the traditional methods by a new distance metric. Furthermore,
we deduce an equivalent MIPS solution to solve the objective of point-to-subspace problem,
which decreases the computational cost significantly. Besides, BSC framework is an asymmetric learning model which handles the complexity of computation and memory as well as the domain differences of videos and images efficiently. 
Extensive experiments on the four datasets, BBT1, UQE50, Vine, FCVID dataset, demonstrated the advantages of our two approaches compared to several state-of-the-arts.

{\small
\bibliographystyle{ieee}
\bibliography{egbib}
}

\end{document}